\documentclass[fleqn,usenatbib]{mnras}

\usepackage{newtxtext,newtxmath}
\usepackage[T1]{fontenc}
\usepackage{ae,aecompl}

\usepackage{graphicx}	% Including figure files
\usepackage{amsmath}	% Advanced maths commands
\usepackage{amssymb}	% Extra maths symbols
\usepackage{float}

\newcommand{\kms}{\mbox{km s$^{-1}$}}
\title[Molecular absorption in Hydra-A]{Deep and narrow CO absorption revealing molecular clouds in the Hydra-A brightest cluster galaxy}
\author[Tom Rose et al.]{Tom Rose,$^{1}$\thanks{E-mail: thomas.d.rose@durham.ac.uk}
A. C. Edge$^{1}$, 
F. Combes$^{2}$,
M. Gaspari$^{3}\thanks{\textit{Einstein} and \textit{Spitzer} Fellow}$,
S. Hamer$^{4}$,
N. Nesvadba$^{5}$,\newauthor
H. Russell$^{4}$,
G. R. Tremblay$^{6}$,
S. A. Baum$^{7,8}$, 
C. O'Dea$^{7,9}$,
A. B. Peck$^{10}$,
C. Sarazin$^{11}$,\newauthor
A. Vantyghem$^{7,12}$,
M. Bremer$^{13}$,
M. Donahue$^{14}$,
A. C. Fabian$^{4}$, 
G. Ferland$^{15}$,\newauthor
B. R. McNamara$^{12}$,
R. Mittal$^{16}$,
J. B. R. Oonk$^{17,18,19}$,
P. Salom\'e$^{2}$,
A. M. Swinbank$^{1}$,\newauthor
M. Voit$^{14}$
\\
$^{1}$Centre for Extragalactic Astronomy, Durham University, DH1 3LE, UK\\
$^{2}$LERMA, Observatoire de Paris, PSL Research Univ., College de France, CNRS, Sorbonne Univ., Paris, France\\
$^{3}$Department of Astrophysical Sciences, 4 Ivy Lane, Princeton University, Princeton, NJ 08544-1001, USA\\
$^{4}$Institute of Astronomy, Cambridge University, Madingly Rd., Cambridge, CB3 0HA, UK\\
$^{5}$Institut d'Astrophysique Spatiale, Centre Universitaire d'Orsay, 91405 Orsay, France\\
$^{6}$Harvard-Smithsonian Center for Astrophysics, 60 Garden St., Cambridge, MA 02138, USA\\
$^{7}$Department of Physics \& Astronomy, University of Manitoba, Winnipeg, MB R3T 2N2, Canada \\
$^{8}$Chester F. Carlson Center for Imaging Science, Rochester Institute of Technology, 84 Lomb Memorial Dr., NY 14623, USA\\
$^{9}$School of Physics and Astronomy, Rochester Institute of Technology, 85 Lomb Memorial Drive, USA\\
$^{10}$Gemini Observatory, Northern Operation Center, 67-0 N. A'Ohoku Place, Hilo, HI, USA \\
$^{11}$Department of Astronomy, 530 McCormick Road, Charlottesville, VA 22904-4325, USA\\
$^{12}$Physics \& Astronomy Department, Waterloo University, 200 University Ave. W., Waterloo, ON, N2L, 2GL, Canada\\
$^{13}$HH Wills Physics Laboratory, Tyndall Avenue, Bristol, BS8 1TL, UK\\
$^{14}$Physics \& Astronomy Department, Michigan State University, East Lansing, MI 48824-2320, USA\\
$^{15}$Department of Physics and Astronomy, University of Kentucky, Lexington, Kentucky 40506-0055, USA\\
$^{16}$RIT College of Science, 85 Lomb Memorial Drive, Rochester, NY 14623, USA\\
$^{17}$SURFsara, P.O. Box 94613, 1090 GP Amsterdam, The Netherlands\\
$^{18}$ASTRON, Netherlands Institute for Radio Astronomy, 7990AA Dwingeloo, The Netherlands\\
$^{19}$Leiden Observatory, Leiden University, Niels Borhweg 2, NL-2333 CA Leiden, The Netherlands\\
}
\date{Accepted XXX. Received YYY; in original form ZZZ}
\pubyear{2018}
\begin{document}
\label{firstpage}
\pagerange{\pageref{firstpage}--\pageref{lastpage}}
\maketitle
\begin{abstract}
Active galactic nuclei play a crucial role in the accretion and ejection of gas in galaxies. Although their outflows are well studied, finding direct evidence of accretion has proved very difficult and has so far been done for very few sources. A promising way to study the significance of cold accretion is by observing the absorption of an active galactic nucleus's extremely bright radio emission by the cold gas lying along the line-of-sight. As such, we present ALMA CO(1-0) and CO(2-1) observations of the \mbox{Hydra-A} brightest cluster galaxy (\mbox{z=0.054}) which reveal the existence of cold, molecular gas clouds along the line-of-sight to the galaxy's extremely bright and compact \mbox{mm-continuum} source. They have apparent motions relative to the central supermassive black hole of between \mbox{-43 and -4 \kms} and are most likely moving along stable, low ellipticity orbits. The identified clouds form part of a \mbox{$\sim$$10^{9}$ $\text{M}_{\odot}$}, approximately edge-on disc of cold molecular gas. With peak CO(2-1) optical depths of \mbox{$\tau$=0.88 $^{+0.06}_{-0.06}$}, they include the narrowest and by far the deepest absorption of this type which has been observed to date in a brightest cluster galaxy. By comparing the relative strengths of the lines for the most strongly absorbing region, we are able to estimate a gas temperature of \mbox{$42^{+25}_{-11}$ K} and line-of-sight  column densities of $N_{ \textnormal{CO}}=2^{+3}_{-1}\times 10 ^{17}\textnormal{cm}^{-2}$ and \mbox{$N_{ \textnormal{H}_{2}  }=7^{+10}_{-4}\times 10 ^{20}\textnormal{cm}^{-2}$}.
\end{abstract}

% Select between one and six entries from the list of approved keywords.
% Don't make up new ones.
\begin{keywords}
\vspace{-0.8cm} clusters: individual: Hydra-A -- radio lines: galaxies -- galaxies: ISM 
\end{keywords}

%%%%%%%%%%%%%%%%%%%%%%%%%%%%%%%%%%%%%%%%%%%%%%%%%%

%%%%%%%%%%%%%%%%% BODY OF PAPER %%%%%%%%%%%%%%%%%%

\section{Introduction}
 
\begin{figure*}
	\includegraphics[width=\textwidth]{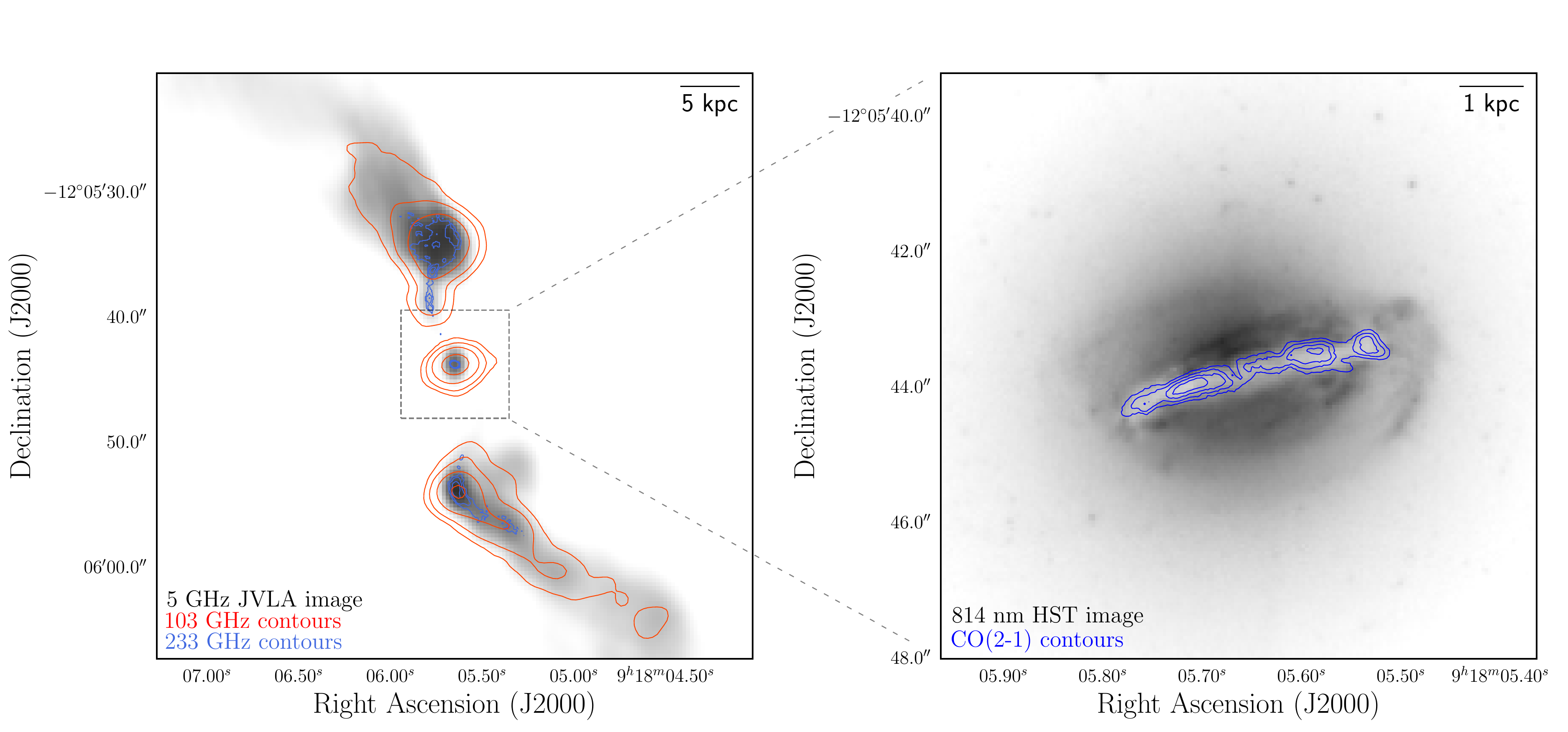}
    \caption{\textbf{Left:} A 5GHz \textit{Karl G. Jansky Very Large Array} (VLA) radio image of Hydra-A showing powerful radio lobes emanating from the centre of the brightest cluster galaxy, with 0.19 arcsec pixel$^{-1}$ resolution (Project 13B-088). Contours of our 233 GHz ALMA continuum image are overlaid in blue and lower angular resolution 103 GHz contours are shown in red. \textbf{Right:} Contours from a CO(2-1) integrated brightness map (described in \S\ref{sec:DiscDynamics}), which mark the 0.15, 0.25, 0.35 and \mbox{0.45 mJy beam$^{-1}$} emission regions, overlaid onto a F814W \textit{Hubble Space Telescope} (HST) near-infrared image of the brightest cluster galaxy \citep{Mittal2015}, with a resolution of \mbox{0.05 arcsec pixel$^{-1}$}. This highlights how the galaxy's dust lane, which is opaque to optical wavelengths, is spatially coincident with a molecular gas disc.}
    \label{fig:HydraAHSTRadioCombinedImage}
\end{figure*}

The accretion of gas onto a galaxy's central supermassive black hole produces powerful outflows of up to \mbox{10$^{46}$ erg s$^{-1}$} which regulate a wide range of physical phenomena, such as gas accretion and star formation \citep{Bower2006,Croton2006,King2015}. Despite being extremely energetic, the processes responsible for the growth of supermassive black holes currently have very limited observational constraints and are far from well understood. For simplicity, many contemporary models and simulations assume that supermassive black holes are fed through highly idealized Bondi accretion. In these models, the supermassive black hole is powered by the smooth and spherical inflow of very hot gas \citep{Bondi1952}, or a similar process based upon a modified version of this approach \citep[e.g.][]{Proga2003, Gaspari2010}. However, models which make use of Bondi accretion do so despite there being a lack of observational evidence indicating that behaviour of this kind is a reality. Instead, recent observations have shown that the central regions of AGN are dominated by the considerably colder molecular gas phase \citep[e.g.][]{Combes2013,GarciaBurillo2014}. Additionally, recent theory and simulations of massive galaxies, groups and clusters point towards supermassive black hole accretion which at sub-parsec scales is chaotic, clumpy and cold, with typical gas temperatures of a few tens of Kelvin \citep[e.g.][]{Pizzolato2005,Pizzolato2010,McNamara2016}. 

In this picture, often described in theoretical works as chaotic cold accretion \citep[or CCA for short, see][]{Pizzolato2005,Gaspari2013,Gaspari2015,Gaspari2017}, the accretion and feedback processes in massive galaxies are linked, at least in part, by radio jets. These jets emanate from a galaxy's central supermassive black hole and inflate buoyant bubbles within the surrounding X-ray gas. In turn, these bubbles then migrate outwards from the galaxy centre. Turbulent motions driven by AGN outflow, jets and bubbles, together with the uplift of low entropy gas material, drive a top-down non-linear condensation of warm filaments and cold clouds. Recurrent inelastic collisions between clouds then promotes the cancellation of angular momentum and boosts the supermassive black hole accretion rate. Behaviour of this kind is expected close to the core, where the velocity dispersion of the molecular gas dominates over the rotational velocity. Subsequent periods of radio-mode AGN feedback are likely to influence further accretion and star formation throughout the galaxy. They are also predicted to significantly affect the thermodynamics of the surrounding intracluster medium gas and to repeatedly shape the evolution of the galaxies within the whole cluster \citep{Churazov2001,Nulsen2005,McNamara2012}. 
%Important observational constraints have been placed on a variety of AGN accretion and feedback models in recent decades, for examples see reviews by \citet{McNamara2007,Fabian2012}. However, further observations are needed to narrow down the properties of the hot and cold gas phases which contribute to the accretion. These properties include the mass, temperature, dynamics and to what extent each phase contributes. Large uncertainties also remain on the fraction of the gas which eventually becomes fuel for future outbursts in the central supermassive black hole and how the different gas phases interact within the cluster. It has been predicted that turbulent mixing between hot and cold gas phases could produce a significant layer of intermediate temperature gas \citep{BegelmanandFabian}. Turbulent mixing could then rapidly cool hot gas within the cooler layers. However, a large number of \mbox{X-ray} observations, \citep[e.g.][]{Peterson2001,Kaastra2001,Sakelliou2002,Peterson2003}, have indicated that less intermediate temperature gas is present within the ICM than previously predicted, kick-starting a now growing consensus that the input of mechanical energy from a central AGN provides the heating required to reconcile an apparent lack of cooling observed in galaxy clusters.

In the last two decades the cooled gas within clusters has been widely surveyed at a range of wavelengths e.g. optical lines \citep{Crawford1999,Tremblay2018}, near infra-red to infra-red \citep{Jaffe2001,Edge2002, Egami2006}, with atomic cooling lines \citep{Edge2010} and with H$_{2}$ observations \citep{Jaffe2005,Donahue2011}. These results show that a significant mass of cold, $<40$ K gas exists within the majority of relaxed galaxy clusters, with typical column densities of $10^{21}-10^{23} \text{ cm}^{-2}$ in the central \mbox{3 - 10 kpc} of the brightest cluster galaxies \citep{Edge2001,SalomeCombes2003}. Important observational constraints can be placed on accretion models such as chaotic cold accretion by determining the properties of this cold gas closer to the supermassive black hole, particularly the typical mass, dynamics and covering fraction. To date, the most direct observational evidence supporting a chaotic cold accretion type model in brightest cluster galaxies has been provided by \cite{David2014} and \cite{Tremblay2016}. Their pencil-beam observations of the \mbox{NGC 5044} and \mbox{Abell 2597} brightest cluster galaxies, respectively, detect cold molecular gas inferred to be within \mbox{$\sim$100 pc} of the galaxy's supermassive black hole. The clouds are detected through absorption by using the AGN as a bright backlight, and give a random snapshot of the properties of the molecular gas which may go on to fuel the AGN. The cold molecular gas regions detected have dynamics consistent with their warmer H\thinspace\small I\normalsize\space clouds and their narrow absorption lines \mbox{(FWHM$\sim$6-30 \kms)} indicate that they may potentially be giant molecular clouds, rather than giant molecular associations, which are falling towards their respective galaxy's supermassive black hole with velocities of between \mbox{240 and 335 \kms}. A larger sample of absorbing clouds is needed to constrain their overall properties, something which is provided by this work with the addition of observations of one more particularly interesting brightest cluster galaxy.

We present new \textit{Atacama Large Millimeter/submillimeter Array} (ALMA) radio observations of Hydra-A (see Fig. \ref{fig:HydraAHSTRadioCombinedImage}), a giant elliptical galaxy with a nearly edge-on disk of dust and cool gas which lies at the centre of an X-ray luminous cluster. Hydra-A is an archetype of a brightest cluster galaxy lying in a cooling flow; it has powerful radio jets and lobes emanating from its centre \citep{Taylor1990}, around which are cavities in the intracluster medium's X-ray emitting gas which have been created by repeated AGN outbursts \citep{McNamara2000,Wise2006}. The radio jets and lobes are almost perpendicular to a rotating gas disc and a spatially coincident dust lane \citep{Hansen1995,Hamer2013}. Hydra-A, or 3C218, is an ideal target for a molecular absorption study for two main reasons. Firstly, it is an extremely strong radio/mm source with amongst the highest flux density in the 3C catalogue of radio sources \citep{Edge1959}. In our ALMA observations, we measure a flux density of $\sim$60 mJy at 230 GHz and in the sample of 35 brightest cluster galaxies observed by \citet{Hogan2015}\footnote{See supplementary data table 2 and 5.}, Hydra-A also has the highest flux density at both 150 and 353 GHz. This makes it an ideal backlight against which to observe molecular absorption. Secondly, previous observations at many different wavelengths suggest that the galaxy's core contains a significant mass of both atomic and molecular gas, making positive detections highly likely e.g. H\thinspace\small{I}\normalsize\space absorption by \citet{Dwarakanath1995,Taylor1996}, CO emission by \citet{Hamer2013} and H$_{2}$ studies by \citet{Edge2002,Donahue2011,Hamer2013}.

Our observations show the rotating disc of molecular gas which was previously observed by \citet{Hamer2013} in much greater detail, thanks to the high angular and spectral resolution which is provided by ALMA. They also reveal a series of cold molecular gas clouds lying along the line-of-sight to the galaxy's AGN. These have apparent velocities of between \mbox{-43 and -4 \kms} relative to the central supermassive black hole and are most likely moving along low ellipticity orbits, with the supermassive black hole at one focus. 

This paper is laid out as follows. In \S\ref{sec:observations} we outline the details of our ALMA observations, in \S\ref{sec:DiscDynamics} we discuss the kinematic properties of Hydra-A's molecular gas disc and in \S\ref{sec:MolecularGasMass} we estimate its mass. The properties of the molecular clouds which lie along the line-of-sight to the mm-continuum source are discussed in \S\ref{sec:accretion} and \S\ref{sec:derivations}. Finally, in \S\ref{sec:discussion} we discuss the significance and implications of our observations. Throughout the paper we assume a flat $\Lambda$CDM Universe with \mbox{$H_{0}=70$ km s$^{-1}$ Mpc$^{-1}$}, \mbox{$\Omega_{M}$=0.3} and \mbox{$\Omega_{\Lambda}$=0.7}. When applying velocity corrections we use a stellar redshift of \mbox{$z$ = 0.0543519}, which provides the best estimate of the gravitational centre of the galaxy. This redshift is calculated from MUSE observations (ID: 094.A-0859) and corresponds to a recession velocity of \mbox{16294 \kms}. This velocity carries an uncertainty of 15 \kms. At Hydra-A's redshift, there is a spatial scale of \mbox{1.056 kpc arcsec$^{-1}$}, meaning that kpc and arcsec scales in figures are approximately equivalent.

\section{Observations}
\label{sec:observations}

\begin{table}
	\centering
	\begin{tabular}{lcr} % four columns, alignment for each

		&  \multicolumn{2}{c}{Observation} \\ 
		\hline
		 & CO(2-1)  & CO(1-0)  \\
		\hline
		Observation date & 2016 Oct 23 & 2018 Jul 18 \\ \\
		Integration time / s & 2700 & 2700 \\ \\
		Velocity resolution / \kms & 1.3 & 2.7 \\ \\
		Frequency resolution / kHz & 977 & 977 \\ \\
		Angular resolution / arcsec & 0.19 & 1.63 \\ \\
		PWV / mm & 1.11 & 2.85 \\ \\
		Field of view / arcsec & 27.8 & 62.1 \\ \\
		Central frequency / GHz & 218.55 & 109.33 \\ \\
		Maximum spacing / km & 1.1 & 0.16 \\ \\
		RMS noise per channel / mJy & 1.4 & 0.7 \\
		\hline 
	\end{tabular}
    \caption{Image properties for the CO(2-1) and CO(1-0) data.}
    \label{tab:observations_table}
\end{table}

The Hydra-A brightest cluster galaxy, located at \mbox{RA 09h18m05.65s}, \mbox{dec. $-12^{\circ}05^{\prime}44.0^{\prime\prime}$ (J2000)}, was observed over two ALMA cycles in bands 3 and 6. For these observations, the ALMA receivers were tuned to emission from the \mbox{$J$=1-0} and \mbox{$J$=2-1} rotational lines of carbon monoxide (CO). This CO emission acts as a tracer for cold molecular hydrogen at temperatures of \mbox{$\lesssim$50 K}. Although molecular hydrogen is significantly more abundant, it is not possible to directly observe it at these temperatures because of a lack of any rotational line emission. 

The spectral line observations from ALMA allowed us to create three dimensional position-velocity maps which trace the brightest cluster galaxy's cold molecular gas. With these high resolution maps, the projected position and line-of-sight velocity of cold molecular gas can be traced throughout Hydra-A.

The first observations taken were during ALMA Cycle 4, of CO(2-1) on 2016 October 23, using band 6. The \mbox{CO(1-0)} observations were taken on 2018 July 18 during ALMA Cycle 5, using band 3. Further technical details of each observation are given in Table \ref{tab:observations_table}.

The data were calibrated using \texttt{CASA} version 5.1.1, a software package which is produced and maintained by the ALMA observatory \citep{CASA}. For each of the observed CO lines, the calibrated data were produced by the ALMA observatory. After their delivery, we performed the necessary continuum subtractions and made channel maps at velocity resolutions to suit our needs.

\section{Dynamics of the cold gas disc}
\label{sec:DiscDynamics}

\begin{figure*}
	\includegraphics[width=\linewidth]{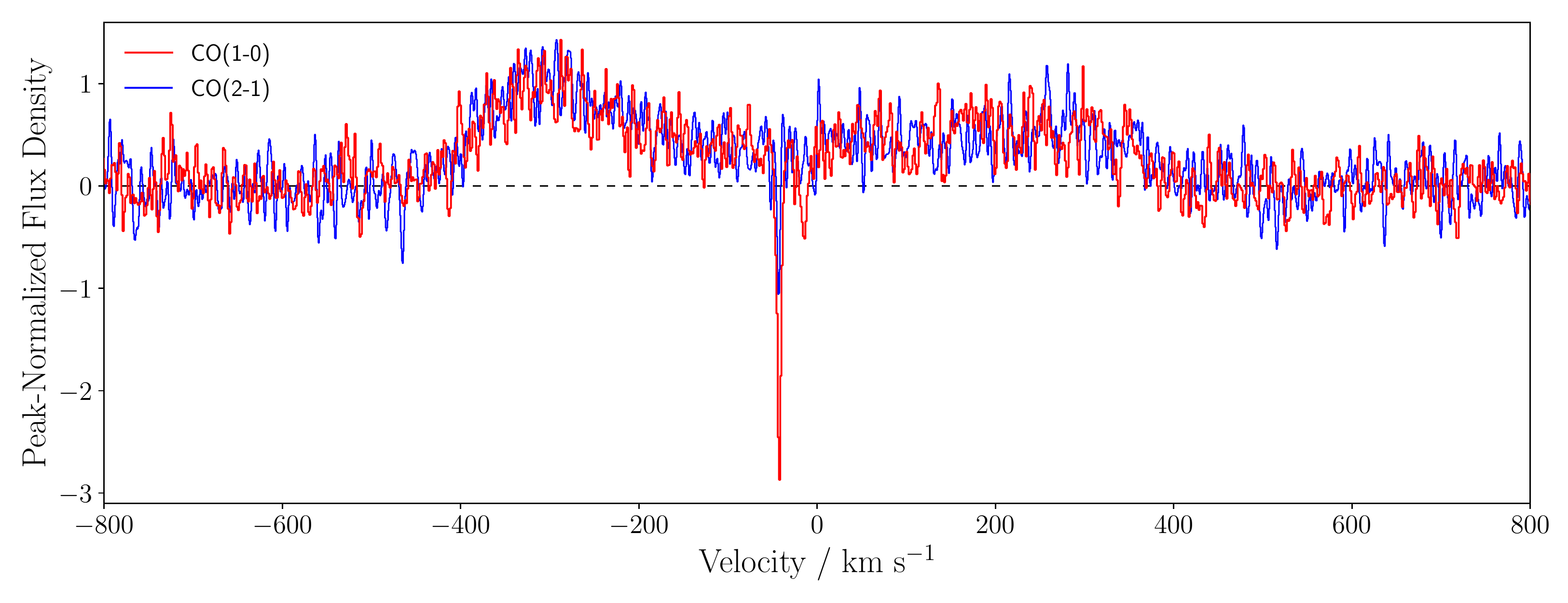}
    \caption{Peak-normalized \mbox{CO(1-0)} and \mbox{CO(2-1)} spectra of Hydra-A's disc from our ALMA observations, extracted using the region indicated by the large blue box shown in Fig. \ref{fig:HydraAContourImage}. The spectra both show a double-peaked structure with a trough at the galaxy's dynamical centre, which itself shows some absorption close to the systemic velocity. The peak-normalization applied to the spectra aids comparison of the emission, but due to the CO(1-0) having a much lower peak, it over emphasizes the strength of the CO(1-0) absorption at the centre. The emission from the CO(2-1) and CO(1-0) continuum-subtracted spectra peak at 13 mJy and 3 mJy, respectively. Additionally, both peak-normalized spectra appear to have a similar noise levels because each observation was done while working towards a similar signal-to-noise ratio. The \mbox{CO(1-0)} and \mbox{CO(2-1)} absorption features are discussed in more detail in \S\ref{sec:accretion} and \S\ref{sec:derivations}.}
    \label{fig:HydraAFullSpectrum}
\end{figure*}

\begin{figure*}
	\includegraphics[width=\linewidth]{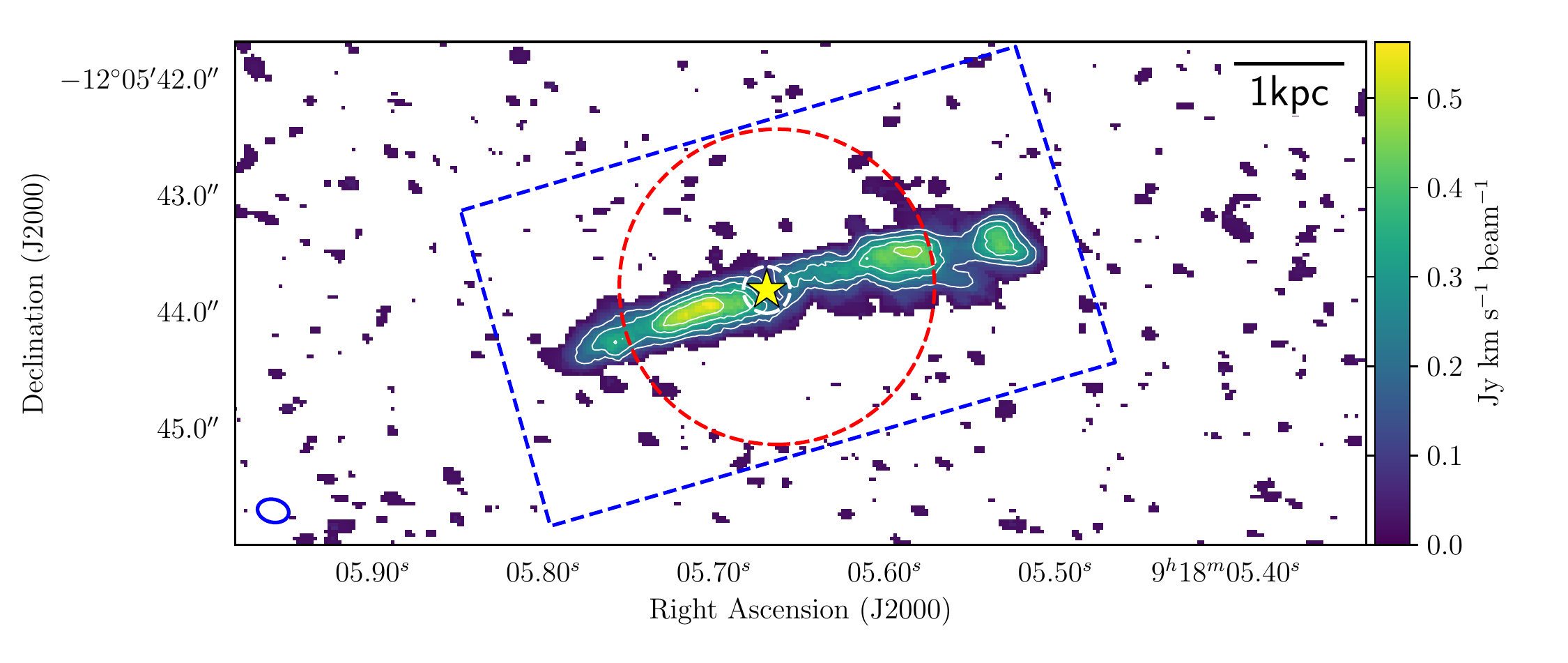}
    \caption{An integrated brightness map of \mbox{${\geq} 3 \sigma$ CO(2-1)} emission in the Hydra-A brightest cluster galaxy. The brightness is integrated from \mbox{-500 km s$^{-1}$} to \mbox{+500 km s$^{-1}$} about the galaxy's systemic velocity. Contours marking 0.15, 0.25, 0.35 and \mbox{0.45 Jy \kms\space beam$^{-1}$} emission are shown in white. The yellow star indicates the position of the mm-continuum source while the large, dashed blue box and dashed white circle indicate the respective regions from which the spectra in Fig. \ref{fig:HydraAFullSpectrum} and \ref{fig:CombinedUncorrectedandCorrectedSpectrum} are extracted. The large, dashed red circle, which is approximately the same size as the CO(1-0) beam, indicates the region from which the spectra in Fig. \ref{fig:1021Spectrum} are extracted. The CO(2-1) beam size is shown in blue in the bottom-left corner.}
    \label{fig:HydraAContourImage}
\end{figure*}

\begin{figure*}
	\includegraphics[width=\linewidth]{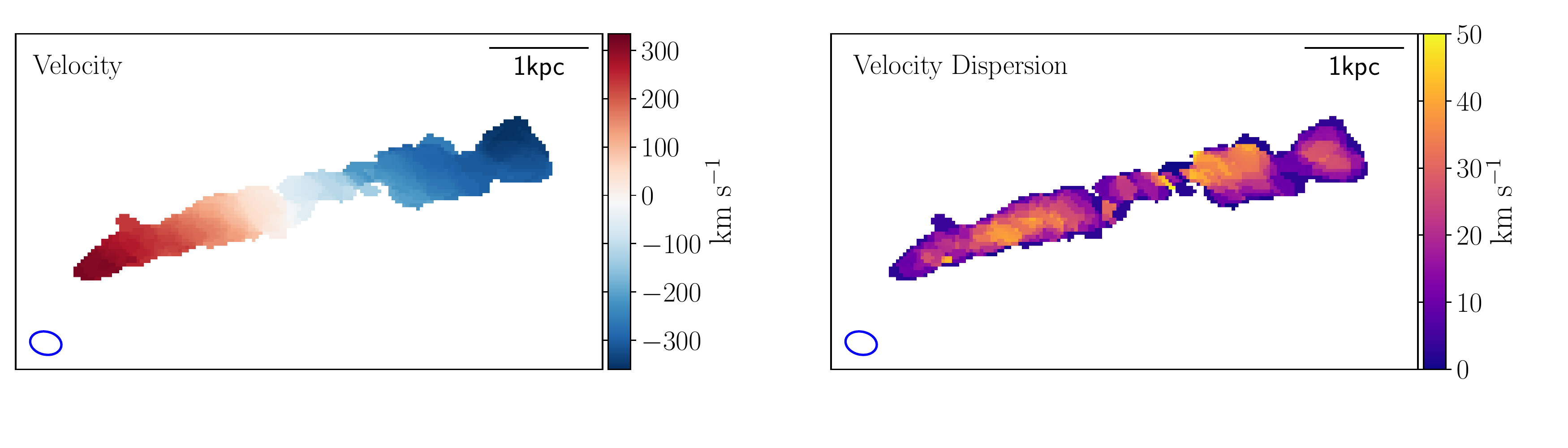}
    \caption{Moments maps of \mbox{${\geq} 5\sigma$} \mbox{CO(2-1)} emission produced using \texttt{CASA}'s \texttt{immoments} task \citep{CASA}. \textbf{Left:} The intensity-weighted velocity of the CO(2-1) line emission relative to the galaxy's systemic recession velocity, showing a disc of cold molecular gas with a line-of-sight velocity gradient of \mbox{${\sim}$140 km s$^{-1}$ kpc$^{-1}$} running across the galaxy's dust lane. \textbf{Right:} The intensity-weighted velocity dispersion of the CO(2-1) emission, which remains significantly smaller than the rotation velocity throughout all but the very centre of the disc. This suggests that the cold molecular gas within Hydra-A's disc is undergoing relatively ordered rotation.}
    \label{fig:CombinedVelocityAndDispersionImage}
\end{figure*}

\begin{figure*}
	\includegraphics[width=\linewidth]{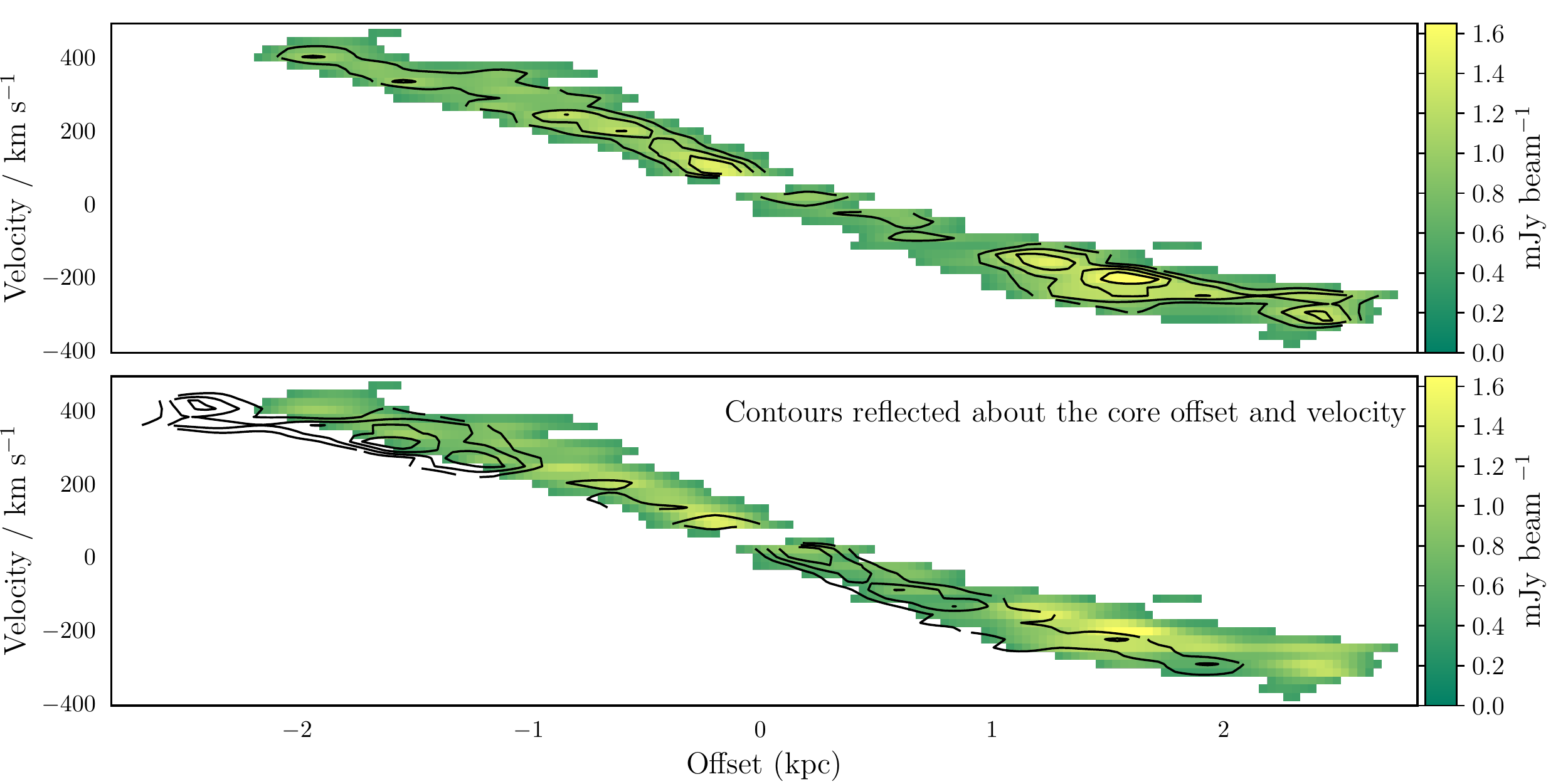}
    \caption{\textbf{Top:} Position-velocity diagram of Hydra-A's ${\geq} 5 \sigma$ CO(2-1) emission along the kinematic axis of the disc, with integrated brightness contours of 0.75, 1.0, 1.25 and 1.5 mJy beam$^{-1}$. The offset is centred at the position of the continuum source and the velocity is centred at Hydra-A's recession velocity. \textbf{Bottom:} Integrated brightness contours reflected about the velocity and offset of the core highlight the slight asymmetry within the disc. Though the disc has a largely symmetric velocity distribution, there is an extended emission region at the extreme edge of the blueshifted side of the disc.}
    \label{fig:Hydra-A_pV_Image}
\end{figure*}

Fig. \ref{fig:HydraAFullSpectrum}\space shows the spectra of the CO(1-0) and CO(2-1) emission observed across Hydra-A's disc. The region from which the spectra are extracted is indicated by the large, blue rectangle on the integrated brightness map shown in Fig. \ref{fig:HydraAContourImage}. Due to its considerably higher angular resolution, we only show an integrated brightness map of the CO(2-1) data, which was produced using the \mbox{\texttt{immoments}} task in \texttt{CASA}. The map reveals a close to edge-on disc of cold molecular gas, with two peaks in brightness on either side of the centre corresponding to those seen in Fig. \ref{fig:HydraAFullSpectrum}. Brightness contours from Fig. \ref{fig:HydraAContourImage} overlaid onto the HST F814 image of Fig. \ref{fig:HydraAHSTRadioCombinedImage} show a cold molecular gas disc that is spatially coincident with the dust lane of the galaxy. The molecular disc and dust lane are aligned with the \mbox{${\sim}$5 kpc} diameter disc of ionized and warm molecular hydrogen identified by \citet{Hamer2013}.

The strong, broad and double-peaked line emission seen in Fig. \ref{fig:HydraAFullSpectrum} results from the disc's rotation and has a large full width zero intensity of \mbox{$\sim$1000 km s$^{-1}$}, consistent with previous CO observations of Hydra-A \citep{ODea1994,Edge2001,Hamer2013}. The double-peaked nature of the CO emission, also previously identified by \citet{Hamer2013}, has so far only been detected in limited number of elliptical galaxies \citep[e.g.][]{Lim2001,Salome2011}, particularly in brightest cluster galaxies. This is due to the low probability of a galactic disc lying almost perfectly edge-on along the line-of-sight. 

Moments maps of the high angular resolution \mbox{CO(2-1)} data, also produced using \texttt{CASA}'s \mbox{\texttt{immoments}} task, show the velocity structure of the disc in greater detail. The intensity-weighted velocity map in the left panel of Fig. \ref{fig:CombinedVelocityAndDispersionImage} shows that the brightest cluster galaxy has an average line-of-sight velocity gradient of \mbox{${\sim}$140 km s$^{-1}$ kpc$^{-1}$}. Throughout the disc the intensity-weighted velocity dispersion shown in the right panel of Fig. \ref{fig:CombinedVelocityAndDispersionImage}, remains small and never exceeds \mbox{$\sim$50 km s$^{-1}$}, implying that the cold gas within the disc is undergoing relatively ordered rotation without large proportions deviating from this significantly. This is further demonstrated by the turbulent Taylor number, the ratio of the rotational velocity to the velocity dispersion. It is clear from Fig. \ref{fig:CombinedVelocityAndDispersionImage} that this ratio is $>1$ throughout the majority of the disc i.e. the rotation velocity dominates over the velocity dispersion. It does however appear to fall below unity in a small region which is coincident with the mm-continuum source; likely because at this point the disc rotation is transverse to the line-of-sight. Chaotic cold accretion simulations \citep[e.g.][]{Gaspari2015,Gaspari2017}, show that where the turbulent Taylor number does reach unity, gas condensation becomes increasingly clumpy \citep[as is seen in the example of NGC 7049, described by][]{Juranova2018}.

Although the velocity dispersion remains small throughout the disc, there are small increases approximately half-way between the galaxy centre and the edges of its visible disc of cold gas which coincide with the peaks in the intensity of the \mbox{CO} emission seen in both Fig. \ref{fig:HydraAFullSpectrum} and \ref{fig:HydraAContourImage}. The increase in the velocity dispersion occurs where the change in the rotational velocity of the disc with respect to the orbital radius is highest, as is demonstrated by the position-velocity diagram shown in Fig. \ref{fig:Hydra-A_pV_Image}. The position-velocity diagram also shows that there are no significant inflows or outflows of cold molecular gas from the disc, as well as highlighting some asymmetry; the blueshifted side of the disc extends to around \mbox{2.5 kpc}, while the redshifted side extends to around \mbox{2 kpc}. 

\section{Molecular Gas Mass of the Disc}
\label{sec:MolecularGasMass}
The total mass of cold molecular gas in Hydra-A can be estimated from both the CO(1-0) and CO(2-1) emission measured along the galaxy's disc using the following relation from \citet{Bolatto2013},

\begin{equation}
\begin{split}
M_{\text{mol}} = \frac{1.05\times 10^{4}}{F_{ul}} \left( \frac{X_{\text{CO}}}{2\times 10^{20}\frac{\text{cm}^{-2}}{\text{K km s}^{-1}}}\right)\left( \frac{1}{1+z}\right) \\ \times \left( \frac{S_{\text{CO}} \Delta v}{\text{Jy km s}^{-1}}\right) \left( \frac{D_{\text{L}}}{\text{Mpc}}\right)^{2} M_{\odot}\enspace ,
\end{split}
\label{eq:massequation}
\end{equation}
where $M_{\text{mol}}$ is the mass of molecular hydrogen, $X_{\text{CO}}$ is the CO-to-H$_{2}$ conversion factor, $z$ is the redshift of the source, $S_{\text{CO}} \Delta v$ is the CO emission integral, $D_{\text{L}}$ is the luminosity distance in Mpc and $M_{\odot}$ is a solar mass. $F_{ul}$ is a factor which is included as an approximate conversion between the expected flux density ratios of the two lines, where $u$ and $l$ represent the upper and lower levels. For CO(1-0), $F_{10}=1$ and for CO(2-1) we use a value of $F_{21}=3.2$. This value is consistent with similar studies \citep[e.g.][]{David2014,Tremblay2016,Tami2018}, and originates from a combination of the factor of two between the frequencies of the lines and the brightness temperature ratio observed for molecular clouds in spiral galaxies of 0.8 \citep{BraineandCombes1992}. For \mbox{Hydra-A}, \mbox{$z=0.0543519$} and \mbox{$D_{\textnormal{L}}=242.4$ Mpc}. We use a standard value of \mbox{$X_{\text{CO}} = 2 \times 10^{20}$ cm$^{-2}$ (K km s$^{-1}$)$^{-1}$} for the CO-to-H$_{2}$ conversion factor in our calculations. This is the same as is used by e.g. \citet{Bolatto2013,David2014,Tremblay2016}. 

To determine the mass of cold molecular hydrogen in Hydra-A's disc and its uncertainties, Monte Carlo simulations which allow us to estimate the emission integral for the spectra shown in Fig. \ref{fig:HydraAFullSpectrum} are performed in the following way. Firstly, for each spectrum the noise in the observed spectrum was estimated from the root mean square (rms) of the continuum emission. This was calculated after excluding the region where the emission is clearly visible (\mbox{-600 \kms} to \mbox{+600 \kms}). Following this, \mbox{100 000} simulated spectra are created based upon the observed spectrum. To produce each simulated spectrum, a Gaussian distribution is created for each velocity channel. This Gaussian distribution is centred at the intensity in the observed spectrum for that particular velocity channel, and has a variance equal to the rms noise squared. A random value for the intensity is drawn from the Gaussian distribution and when this has been done across all velocity channels, a simulated spectrum is produced. A double-Gaussian line provides the best fit and for each of the \mbox{100 000} spectra produced such a fit is found using $\chi^{2}$ minimisation. Finally, the molecular mass is found by using Eq. \ref{eq:massequation} and the median value resulting from the \mbox{100 000} estimates of $S_{\text{CO}} \Delta v$. The upper and lower ${1\sigma}$ errors are taken from the values which delimit the 15.865 per cent highest and lowest results for each of the fits (i.e. 68.27 per cent of the fitted parameters will therefore lie within this ${1\sigma}$ range). This procedure treats the CO(1-0) and CO(2-1) spectra independently and provides two inferred masses for Hydra-A's cold molecular gas. Using the CO(1-0), we find an estimated mass of \mbox{$M_{\text{mol}}=1.06^{+0.02}_{-0.02}\times 10^{9}$ $\text{M}_{\odot}$}, while for the CO(2-1), we estimate a mass of \mbox{$M_{\text{mol}}=5.6^{+0.1}_{-0.1}\times 10^{9}$ $M_{\odot}$}. It is important to note that the errors quoted in these values are due to the noise seen in each of the spectra. In reality, the molecular mass may differ from the estimated values by as much as an order of magnitude due to the uncertainty surrounding the values of the X-factor and the conversion factor, $F_{ul}$. However, given these larger errors, the masses we find are consistent with the value of \mbox{$M_{\text{mol}}=2.26 \pm 0.29\times 10^{9}$ $M_{\odot}$} found from \textit{Institut de RadioAstronomie Millim\'etrique} (IRAM) 30m observations observations presented in \citet{Hamer2013} and are typical molecular gas masses when compared with those found through CO detections for other brightest cluster galaxies with cooling flows \citep[e.g.][]{Edge2001,SalomeCombes2003}.

\section{Cold, Clumpy Gas Clouds}
\label{sec:accretion}

\begin{figure*}
	\includegraphics[width=\textwidth]{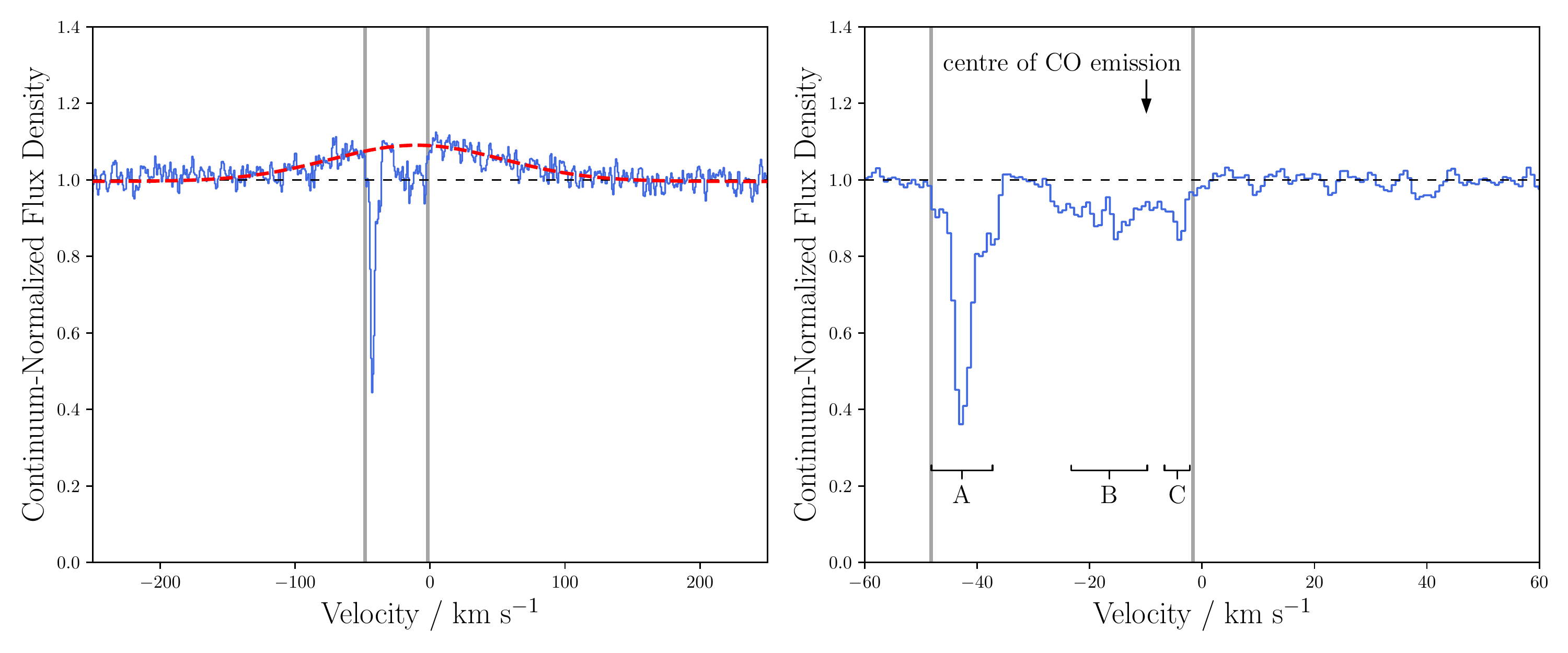}
    \caption{\textbf{Left:} Hydra-A's continuum-normalized \mbox{CO(2-1)} spectrum, extracted from the region lying along the line-of-sight to the bright mm-continuum source (the exact region is shown by a dashed, white circle in Fig. \ref{fig:HydraAContourImage}). The spectrum shows a broad emission component \mbox{(FWHM=151$^{+8}_{-7}$ \kms)} and very clear absorption regions. The broad component is present throughout the disc and is produced by emission from a large scale ensemble of cold molecular gas clouds. To show the strength of the absorption regions more clearly, a Gaussian line is fitted to the spectrum (shown in red). This fit is made after excluding the spectral bins where absorption is visible (approximately \mbox{-48 \kms} to \mbox{-1 \kms}, indicated by the vertical grey lines). \textbf{Right:} The continuum-normalized \mbox{CO(2-1)} spectrum extracted from the region lying along the line-of-sight to the mm-continuum source, with the broad emission component removed. The absorption features are due to cold molecular gas clouds which are likely to be within $\sim$200 pc of the supermassive black hole. The best fit parameters of the absorbing regions labelled `A', `B' and `C' are shown in Table \ref{tab:results_table}.}
    \label{fig:CombinedUncorrectedandCorrectedSpectrum}
\end{figure*}

\begin{table*}
	\centering
	\begin{tabular}{lcccr} % four columns, alignment for each
		\hline
		Region & $v_{\text{cen}}$ / \kms & $v_{\text{cen}}$ - $v_{\textnormal{emission}}$ / \kms & FWHM / \kms & $\tau_{\text{max}}$\\
        \hline
        Emission & -10 $^{+2}_{-2}$ &- & 151$^{+8}_{-7}$ & - \\ \\
		A & -42.8 $^{+0.1}_{-0.1}$ & -33$^{+2}_{-2}$& 4.4 $^{+0.3}_{-0.3}$ & 0.88$^{+0.06}_{-0.06}$\\ \\
		B & -16.5 $^{+0.8}_{-0.8}$ & -7$^{+2}_{-2}$ & 17 $^{+2}_{-2}$ & 0.11 $^{+0.01}_{-0.01}$\\ \\
        C & -4.4 $^{+0.2}_{-0.3}$&+6$^{+2}_{-2}$ & 2.3$^{+1.6}_{-0.8}$ & 0.13 $^{+0.03}_{-0.03}$\\ 
		\hline
	\end{tabular}
    \caption{Best fit parameters for the broad emission and absorption regions along the line of sight to Hydra-A's mm-continuum source, as seen in Fig. \ref{fig:CombinedUncorrectedandCorrectedSpectrum}. The 1$\sigma$ errors in the v$_{\textnormal{cen}}$ values are in addition to the systemic uncertainty of \mbox{15 \kms} in the recession velocity of Hydra-A taken from MUSE observations of stellar absorption lines. We also include the velocities of the absorption regions relative to the centre of the CO(2-1) line-of-sight emission, since this provides a second plausible value for the zero velocity reference point.}
    \label{tab:results_table}
\end{table*}

\begin{figure}
	\includegraphics[width=\columnwidth]{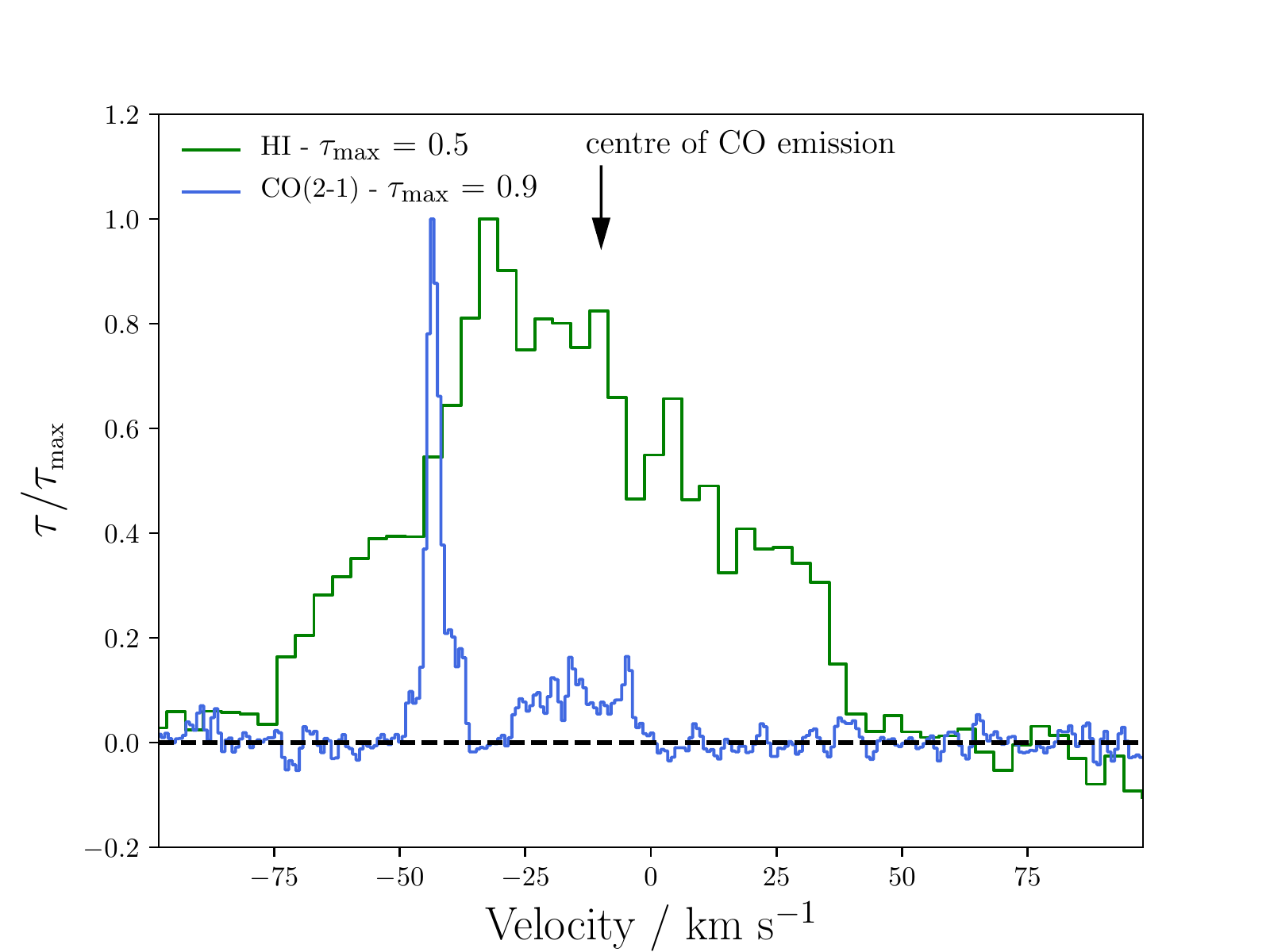}
    \caption{Hydra-A's normalized CO(2-1) and H\thinspace\small{I}\normalsize\space optical depth profiles for the region co-spatial with the mm-continuum source. The CO(2-1) optical depth profile is produced using the spectrum shown in the the right panel of Fig. \ref{fig:CombinedUncorrectedandCorrectedSpectrum}, and the H\thinspace\small{I}\normalsize\space profile is taken from VLBA observations presented in \citet{Taylor1996}.}
    \label{fig:ApparentOpticalDepth}
\end{figure}

As can be seen in Fig. \ref{fig:HydraAFullSpectrum}, the \mbox{CO(1-0)} and \mbox{CO(2-1)} spectra from Hydra-A's disc show the presence of a narrow absorption feature close to the zero-velocity point, caused by molecular clouds which lie along the line-of-sight to the bright mm-continuum source. Due to the high angular resolution of the CO(2-1) observations, this absorption can be seen in more detail by extracting spectra from a much smaller region. The left panel of Fig. \ref{fig:CombinedUncorrectedandCorrectedSpectrum} shows the continuum-normalized \mbox{CO(2-1)} spectrum along the line-of-sight to the mm-continuum source, taken from the region shown by the small, dashed white circle in Fig. \ref{fig:HydraAContourImage}. The spectrum, which is centred on the stellar recession velocity of the galaxy, has two main features. Firstly, there is a broad emission component. This can also be seen across the rest of the galaxy's disc and is produced by a large-scale ensemble of molecular gas clouds. It is this component which, when summed across the disc, produces the double-peaked spectrum seen in Fig. \ref{fig:HydraAFullSpectrum}. Secondly, the spectrum reveals a group of absorption features which are produced by the cold molecular gas clouds which lie along the line-of-sight to the bright and unresolved continuum source. Importantly, the compact nature of the continuum source ensures that we are probing a single line-of-sight and seeing no contamination from more extended emission. These absorption features appear to be unique to the radio core of the galaxy and do not arise elsewhere, such as along the line-of-sight to the bright spots in galaxy's radio lobes.

To show the strength of the absorption more clearly, the broad emission component is removed by fitting and subtracting a Gaussian line to the spectrum, with the Gaussian fit excluding the spectral bins where absorption can be seen (approximately \mbox{-48 to -1 \kms}, as indicated by the vertical grey lines in Fig. \ref{fig:CombinedUncorrectedandCorrectedSpectrum}). This region was chosen by performing Gaussian fits to the emission after applying masks to the spectrum with limits at every spectral bin between \mbox{-55 $\pm$ 10 \kms} and \mbox{+5 $\pm$ 10 \kms}. The selected range produces a spectrum with the lowest \mbox{$\chi^{2}_{\nu}$} value when the non-masked region is fitted to a flat line after the removal of the emission. The right panel of Fig. \ref{fig:CombinedUncorrectedandCorrectedSpectrum} shows the spectrum after the subtraction of the optimum Gaussian fit. 

Following this correction, an apparent optical depth profile, $\tau_{\text{a}}$, was calculated from the observed intensities using the relation \citep{Savage1991},
\begin{equation}
\tau_{\text{a}} = \ln{ [ I_{\text{0}}/I_{\text{obs}}(\lambda)]}\text{\space},
\end{equation}
where $I_{\text{0}}$ is the continuum intensity (independent of $\lambda$ over the wavelength range observed) and $I_{\text{obs}}(\lambda)$ is the observed intensity. Due to the high signal to noise ratio and because the spectrum's velocity resolution is relatively high compared with the line widths, our calculations of $\tau_{\text{a}}$ are likely to be minimally affected by the instrumental spread function. In this case $\tau_{\text{a}}$ is a good representation of the optical depth, $\tau$, and so we treat the two as being equivalent. 

We find that a triple-Gaussian line provides the best fit to the absorption features seen in Fig. \ref{fig:CombinedUncorrectedandCorrectedSpectrum}. As such, we estimate the properties of these emission lines using an appropriately modified version of the Monte Carlo method previously described in \S\ref{sec:MolecularGasMass}. We assume a filling factor of 1, so our calculations provide a lower limit on the optical depths of the absorbing regions. The estimates for the optical depths of the absorbing regions are shown in Table \ref{tab:results_table}, along with their velocities, $v_\textnormal{cen}$, and the FWHM of the lines. As well as the observed MUSE stellar redshift the spectrum is centred on, we show the velocities of the absorption regions relative to the line-of-sight emission since this provides a second plausible value for the zero velocity reference point.

One particularly strong absorption feature (labelled `A' in Fig. \ref{fig:CombinedUncorrectedandCorrectedSpectrum}) can be seen in the spectrum, where a peak of around 60 per cent of the continuum emission is absorbed. This has a narrow line width of \mbox{FWHM=4.4$ ^{+0.2}_{-0.3}$ km s$^{-1}$}; the narrowest such absorption feature found to date in a brightest cluster galaxy. Previous observations by \citet{David2014,Tremblay2016} have comparable velocity resolution, so this is unlikely to be the narrowest feature observed purely because their absorption regions are not sufficiently resolved in velocity. This absorption region is found to have a high peak optical depth of \mbox{$\tau$=0.88$^{+0.06}_{-0.06}$}. Therefore, it is also by far the deepest CO absorption line detected so far in a brightest cluster galaxy, with the next strongest being found in \mbox{Abell 2597} with an optical depth of \mbox{${\tau \sim}$0.3} \citep{Tremblay2016}. At the line's peak around \mbox{${\sim}$60} per cent of the emission from the continuum source is being absorbed, either by optically thick gas which covers \mbox{${\sim}$60} per cent of the continuum source spatially, or by more diffuse gas which completely covers the source's synthesised beam with an optical depth of \mbox{${\tau \sim}$0.9}. Further absorption regions due to clouds moving away from the mm-continuum source at slightly lower velocities are also present. The central velocity, FWHM and peak optical depth of each of these absorption features are also shown in Table \ref{tab:results_table}.

Collectively, these absorption features reveal the presence of cold molecular gas clouds moving relative to the galaxy's central supermassive black hole at velocities of between  \mbox{-43} and \mbox{-4 km s$^{-1}$}. However, these are simply apparent motions along the line-of-sight and do not necessarily imply that the gas clouds are flowing away from the galaxy centre at those velocities. Nevertheless, simulations predict that over long periods of time clouds such as those observed here experience collisions in the inner region which reduce their angular momentum. This disrupts their relatively stable, slightly elliptical orbits and funnels them towards the galaxy's supermassive black hole \citep{Gaspari2017, Gaspari2018}. However, it is also possible that the slightly blueshifted velocities seen could be due to a bar, though this is typically associated with lower mass spiral galaxies.

As well as cold molecular gas, warmer atomic gas has also been detected in Hydra-A with the \textit{Very Long Baseline Array} (VLBA) telescope \citep{Taylor1996}. These observations detect H\thinspace\small I\normalsize\space gas in absorption along multiple lines-of-sight close to the centre of the galaxy; absorption is seen against the knots of the active galactic nucleus's jets, as well as against the extremely bright continuum source. 

The optical depth profiles of the H\thinspace\small I\normalsize\space and CO(2-1) absorption for the line-of-sight to the continuum source are shown in Fig. \ref{fig:ApparentOpticalDepth}, where both have been normalized to aid comparison. Both optical depth profiles probe the region along the line-of-sight to the continuum source. The H\thinspace\small{I}\normalsize\space line lies at 1.4 GHz and the CO(2-1) line lies at \mbox{230 GHz}. Due to the large difference in these frequencies, if there are spectral index variations or synchrotron opacity effects in the radio core, it is possible that different parts of the core would dominate at the different frequencies. However, in both cases the absorption is observed against an unresolved core and so this should not be an issue when comparing the two spectra. 

Both the warm atomic gas and cold molecular gas appear to be centred at slightly blueshifted velocities relative to the supermassive black hole. Additionally, both absorption profiles have velocity ranges with comparable widths and profiles which may suggest that the clouds detected as a narrow CO(2-1) line are visible as a blended H\thinspace\small I\normalsize\space line. Given that the clouds are spatially coincident, if this is the case it would be consistent with the hypothesis that when heated, cold molecular gas regions produce an encompassing skin of warm atomic gas \citep[e.g.][]{Jaffe2005}. With the exception of the small velocity discrepancy between the peaks in the optical depths of the atomic and molecular gas, this relationship is similar to that seen in Abell 2597 \citep{Tremblay2016} and demonstrates the need for an increase in the number of joint CO and H\thinspace\small I\normalsize\space observations to advance our understanding of the link between the atomic and molecular gas and how they fit into the predictions of accretion models and simulations. Nevertheless, the spatially coincident detections of clumpy molecular and atomic gas clouds in these systems provides a strong argument against models related to Bondi accretion, which rely upon the smooth and spherically symmetric accretion of hot gas. The detections made so far are more consistent with the predictions of clumpy accretion models such as chaotic cold accretion.

\section{Temperature and Column Density Derivations}
\label{sec:derivations}

\begin{figure}
	\includegraphics[width=\columnwidth]{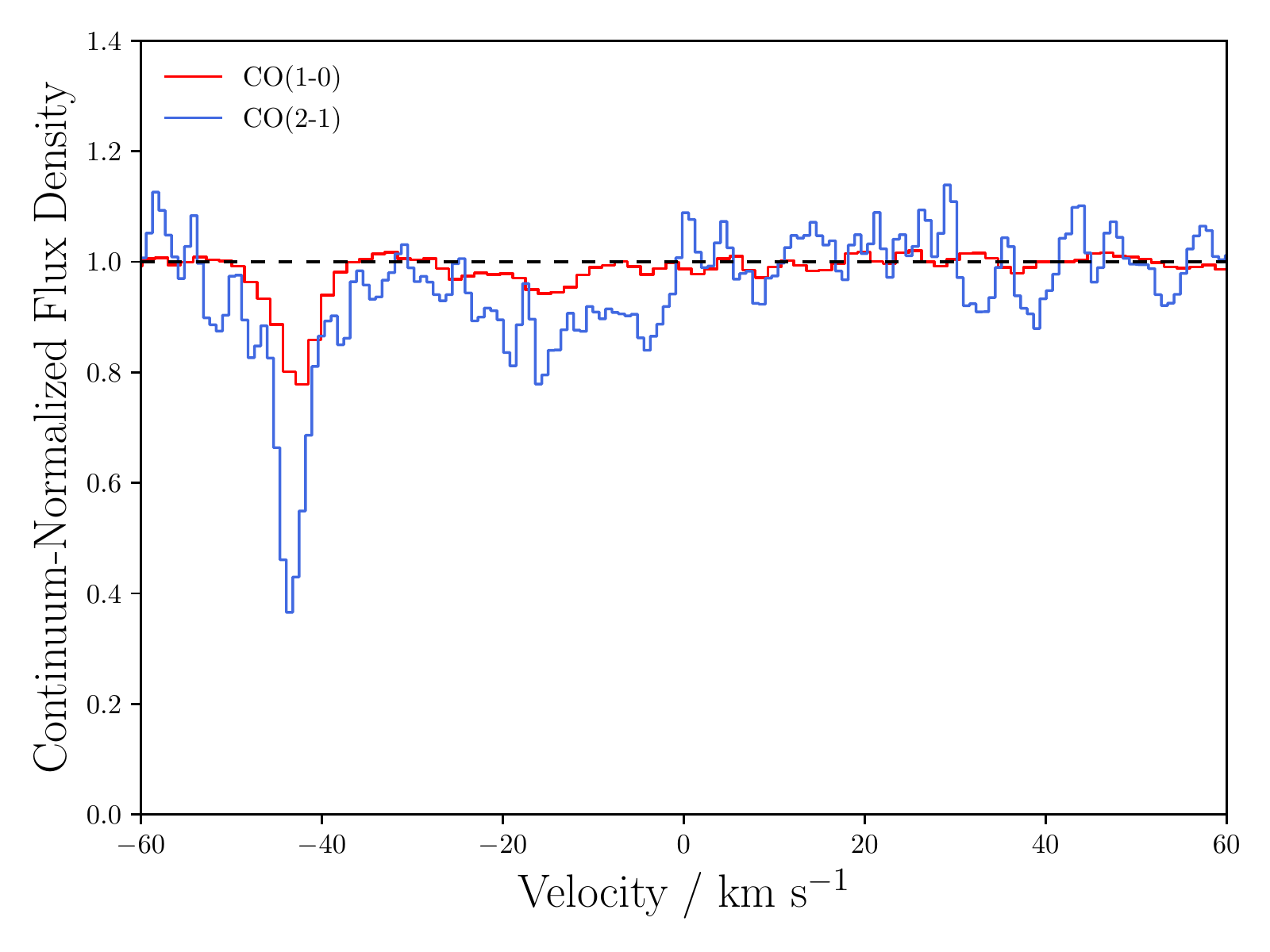}
    \caption{Hydra-A's continuum-normalized CO(1-0) and \mbox{CO(2-1)} flux density after the removal of the emission which is seen across the galaxy's disc. These spectra are both extracted from the region indicated by the large, dashed red circle in Fig \ref{fig:HydraAContourImage}. A wider region is used to extract these spectra due to the larger beam size of the CO(1-0) images. By comparing the strength of the absorption seen for the two lines, we find that the absorbing gas has a temperature of $T_\textnormal{ex} = 42^{+25}_{-11}$ K and line-of-sight column densities of $N_{ \textnormal{CO} }=2^{+3}_{-1}\times 10 ^{17}\textnormal{cm}^{-2}$ and \mbox{$N_{ \textnormal{H}_{2}  }=7^{+10}_{-4}\times 10 ^{20}\textnormal{cm}^{-2}$}.}
    \label{fig:1021Spectrum}
\end{figure}

Although the angular resolution of our CO(2-1) observations is considerably higher than those of CO(1-0), both have sufficient spectral resolution to show the narrow absorption along the line-of-sight to the mm-continuum source. Fig. \ref{fig:1021Spectrum} shows the spectra for each CO line, extracted from the region shown by the large, dashed red circle in Fig. \ref{fig:HydraAContourImage}, which is of a similar size to the CO(1-0) beam. 

The strength of the CO(1-0) absorption line is primarily dependent on the number of CO molecules in the ground rotational state ($J$=0), whereas the absorption line observed in CO(2-1) primarily depends on the number of molecules in the first rotational state ($J$=1). The relative number of molecules in these two states is determined by the temperature of the gas. Therefore, the ratio of the optical depth values for these two absorption lines will give a direct measure of the gas temperature, assuming that the lines are not optically thick. Due to the lower resolution and larger beam size of the CO(1-0) observations, only the most strongly absorbing region remains clearly visible in both spectra when the larger extraction region is used, so we are only able to estimate the temperature for this absorption region.

For optically thin gas in local thermodynamic equilibrium, the \mbox{CO(1-0)} and \mbox{CO(2-1)} velocity integrated optical depths are related by,
\begin{equation}
\label{eq:opacityratio}
\frac{\int \tau_{21} dv}{\int \tau_{10} dv} = 2 \frac{1 - e^{- h\nu_{21}/k T_{\textnormal{ex}}}}{e^{h\nu_{10}/k T_{\textnormal{ex}}} -1}\enspace ,
\end{equation}
where $h$ and $k$ are the Planck and Boltzmann constants, $\nu_{10}$ and $\nu_{21}$ are the rest frequencies of the \mbox{CO(1-0)} and \mbox{CO(2-1)} lines and $T_{\textnormal{ex}}$ is the excitation temperature \citep{Bolatto2003,Godard2010,Magnum2015}. In the above, CO is approximated to being a stiff molecule such that $\nu_{10}$ and $\nu_{21}$ differ by a factor of two. To estimate $T_{\textnormal{ex}}$ for the strongly absorbing cloud, we carry out calculations using a Monte Carlo approach similar the one described in \S\ref{sec:MolecularGasMass}, again using 100 000 simulated spectra for each line. For each pair of simulated spectra, the emission is removed and the optical depth integrals are estimated using a single Gaussian fit to each line. We then use Eq. \ref{eq:opacityratio} to calculate $T_\textnormal{ex}$ numerically and find a cloud excitation temperature of \mbox{$T_\textnormal{ex} = 42^{+25}_{-11}$ K}.

Using the derived excitation temperature, it is also possible to estimate the total column density, $N_{\textnormal{tot}}$, of the molecular cloud. In general,
\begin{equation}
\label{}
N_{\textnormal{tot}} = Q(T_{\textnormal{ex}}) \frac{8 \pi \nu_{ul}^{3}}{c^{3}}\frac{g_{l}}{g_{u}}\frac{1}{A_{ul}} \frac{1}{ 1 - e^{-h\nu_{ul}/k T_{\textnormal{ex}}}}\int \tau_{ul} dv \enspace ,
\end{equation}
where $Q$($T_{\textnormal{ex}}$) is the partition function, $c$ is the speed of light, $A_{ul}$ is the Einstein coefficient of the observed transition and $g$ the level degeneracy, with the subscripts $u$ and $l$ representing the upper and lower levels \citep{Godard2010,Magnum2015}. Using the derived excitation temperature and the velocity integrated optical depth of the CO(2-1) line, we find a line-of-sight CO column density of $N_{ \textnormal{CO}}=2^{+3}_{-1}\times 10 ^{17}\textnormal{cm}^{-2}$; a similar CO column density to the absorption regions identified by e.g. \citep{Wiklind2018}. This value can also be used to estimate the column density of molecular hydrogen by assuming a carbon abundance equal to the Milky Way gas phase, and assuming that all gas-phase carbon exists in CO molecules \citep{Sofia}. In this case, there is a conversion factor of CO/H$_{2}=3.2\times 10^{-4}$, giving an H$_{2}$ column density of \mbox{$N_{ \textnormal{H}_{2}  }=7^{+10}_{-4}\times 10 ^{20}\textnormal{cm}^{-2}$}. 

Observations of Hydra-A's X-ray emission by \citet{Russell2013} have also been used to estimate the total H$_{2}$ column density along the line-of-sight to the core. They find a value of $N_{ \textnormal{H}_{2}} = 3.5\pm 0.3 \times 10^{22}$ cm$^{-2}$, many times larger than our estimated value. However, this difference is likely due to the fact that our estimate is for the mass of gas which lies in the most strongly absorbing region. The \mbox{X-ray} observations of \citet{Russell2013} would instead probe all of the gas along this line-of-sight.

\section{Discussion}
\label{sec:discussion}

The near edge-on orientation of Hydra-A's dust and gas discs is a rarity and so our observations of this brightest cluster galaxy provide a unique opportunity to analyse the dynamics of its molecular gas disc with crucial knowledge of how it is placed within the overall structure of the galaxy. The principal absorption line detected in Hydra-A is by far the deepest found to date in a brightest cluster galaxy, with a peak CO(2-1) optical depth of \mbox{${\tau} = 0.88^{+0.06}_{-0.06}$}, a temperature of \mbox{$T_\textnormal{ex} = 42^{+25}_{-11}$ K} and line-of-sight column densities of \mbox{$N_{ \textnormal{CO}}=2^{+3}_{-1}\times 10 ^{17}\textnormal{cm}^{-2}$} and \mbox{\mbox{$N_{ \textnormal{H}_{2}  }=7^{+10}_{-4}\times 10 ^{20}\textnormal{cm}^{-2}$}}. The strength of this absorption is likely due to the edge-on nature of the disc. As a result of this orientation, the column density of the molecular gas intersecting the line-of-sight to the mm-continuum source is particularly high and so very strong absorption is seen.

Although it is not possible to directly measure the velocity component of the molecular gas clouds orthogonal to the line-of-sight, the \mbox{${\sim}$140 \kms\space kpc$^{-1}$} line-of-sight velocity gradient across the disc of the galaxy implies that unless they lie at very large radii, they should be significantly larger than the infall velocities of the molecular clouds. However, it is unlikely that the clouds we detect lie at large distances from the supermassive black hole. Simulations by \citet{Gaspari2017} of clumpy molecular gas condensation show that both the volume filling factor and internal density of molecular clouds are inversely proportional to radius. Therefore, most of the dense, absorbing clouds contributing to the line-of-sight obscuration are expected to reside in the inner region, within radii of up to $\sim$200 pc. Conversely, the clouds' low velocities relative to the supermassive black hole suggest that they lie beyond the Bondi capture radius of \mbox{$\sim$10 pc}, within which they would gain large velocities towards the core due to the increasing likelihood of collisions and the strength of the supermassive black hole's gravitational field. 

Regarding the velocity structure of the disc, the large inferred horizontal velocities, combined with the comparatively small line-of-sight velocities of a few tens of \kms\space away from the core, suggest the molecular clouds are moving within a relatively stable, slightly elliptical disc. The asymmetry seen in Hydra-A's position-velocity diagram also point towards a low ellipticity disc. However, it is also possible that the non-circular motion inferred in the absorbing clouds may be caused by a bar, though this would typically be associated with spiral galaxies of lower mass. %Given that the more extended edge of the disc is on the blueshifted side and that the absorbing clouds and the CO emission are blueshifted, the mm-continuum source will be at the less proximate focus of an anti-clockwise rotating disc or the more proximate focus of a clockwise rotating disc.

In terms of the line-of-sight velocities of the molecular clouds, this observation is in contrast to those of \citet{David2014} and \citet{Tremblay2016}, where velocities of \mbox{$\sim 200 - 300$ \kms} towards the supermassive black hole were observed. Additionally, in Hydra-A's case the absorbing clouds along the line-of-sight to the core lie roughly at the centre of the broad CO emission (shown in the left panel of Fig. \ref{fig:CombinedUncorrectedandCorrectedSpectrum}). This implies that clouds are not yet falling within the black hole's influence region ($< 10$ pc), but drifting in the turbulent field at up to $\sim 200$ pc. Indeed, our pencil-beam observations are simply showing a random sample of molecular clouds which correlate well with the line-of-sight velocity dispersion close to the core.% and supports the argument that the molecular gas disc in Hydra-A is slightly elliptical, with the mm-continuum source at one focus.

This work represents one of very few detections of cold molecular gas lying in front of a brightest cluster galaxy's strong and unresolved mm-continuum source. As such, it is difficult to make more general predictions regarding the properties of the molecular gas that would be expected in other brightest cluster galaxies. A larger, more statistically significant survey is needed to investigate the universal properties of this molecular gas, such as the covering fraction of cold molecular gas clouds, their temperatures and masses.

Further ALMA cycle 5 \mbox{CO(1-0)} observations of a significantly larger sample of 21 extremely bright, core dominated radio sources with unresolved emission at \mbox{${85 - 110}$ GHz} of \mbox{${>}10$ mJy} and which have been drawn from an X-ray selected sample of over 1000 clusters of galaxies will be presented in Rose et al. (in preparation). Further scheduled Cycle 6 ALMA observations of Hydra-A will also allow us to examine the properties of the galaxy's molecular gas in much greater detail. We will be able to place constraints on the isotopic ratios of carbon and oxygen using \mbox{$^{13}$CO(1-0)}, \mbox{$^{13}$CO(2-1)}, \mbox{C$^{18}$O(1-0)}, \mbox{C$^{18}$O(2-1)} and \mbox{C$^{17}$O(1-0)} lines. Observations of \mbox{HCN(1-0)}, \mbox{HNC(1-0)}, \mbox{HCO+(1-0)}, \mbox{CS(2-1)} and \mbox{CN(2-1)} will allow us to determine the column density of these molecules relative to CO in the absorbing clouds. The study will also provide additional constraints on the temperature and column densities of the molecular gas along the line-of-sight.

\section*{Acknowledgements}

T.R. is supported by the Science and Technology Facilities Council (STFC) through grant ST/R504725/1.

A.C.E. acknowledges support from STFC grant ST/P00541/1.

The work of S.B. and C.O. was supported by NSERC (Natural Sciences and Engineering Research Council of Canada).

M.G. is supported by NASA through Einstein Postdoctoral Fellowship Award Number PF5-160137 issued by the Chandra X-ray Observatory Center, which is operated by the SAO for and on behalf of NASA under contract NAS8-03060. Support for this work was also provided by Chandra grant GO7-18121X.

G.R.T. acknowledges support from the National Aeronautics and Space Administration (NASA) through Chandra Award Number GO7-8128X8, issued by the Chandra X-ray Center, which is operated by the Smithsonian Astrophysical Observatory for and on behalf of NASA under contract NAS8-03060.

This paper makes use of the following ALMA data: ADS/JAO.ALMA\#2016.1.01214.S, ADS/JAO.ALMA\#2017.1.00629.S. ALMA is a partnership of ESO (representing its member states), NSF (USA) and NINS (Japan), together with NRC (Canada) and NSC and ASIAA (Taiwan) and KASI (Republic of Korea), in cooperation with the Republic of Chile. The Joint ALMA Observatory is operated by ESO, AUI/NRAO and NAOJ.

This work used the DiRAC Data Centric system at Durham University, operated by the Institute for Computational Cosmology on behalf of the STFC DiRAC HPC Facility (www.dirac.ac.uk). This equipment was funded by BIS National E-infrastructure capital grant ST/K00042X/1, STFC capital grant ST/H008519/1, and STFC DiRAC Operations grant ST/K003267/1 and Durham University. DiRAC is part of the National E-Infrastructure.

This research made use of APLpy, an open-source plotting package for Python \citep{aplpy}.

The authors gratefully acknowledge the referee Dr. Rick Perley for his comments, which helped us to improve the paper. We give him extra thanks for providing the JVLA dataset used in Fig. \ref{fig:HydraAHSTRadioCombinedImage}.

%%%%%%%%%%%%%%%%%%%%%%%%%%%%%%%%%%%%%%%%%%%%%%%%%%

%%%%%%%%%%%%%%%%%%%% REFERENCES %%%%%%%%%%%%%%%%%%

% The best way to enter references is to use BibTeX:

\bibliographystyle{mnras}
\bibliography{bibliography} % if your bibtex file is called example.bib

% Don't change these lines
\bsp	% typesetting comment
\label{lastpage}
\end{document}